\documentclass[12pt,epsf,epsfig,psfig]{article}
\usepackage{graphicx}
\usepackage{epsfig}
\def\e{\epsilon}

\def\be{\begin{equation}}
\def\ee{\end{equation}}

\def\lsim{\raise0.3ex\hbox{$<$\kern-0.75em\raise-1.1ex\hbox{$\sim$}}}
\def\gsim{\raise0.3ex\hbox{$>$\kern-0.75em\raise-1.1ex\hbox{$\sim$}}}

\def\NP{{ Nucl.\ Phys.\ }}
\def\PL{{ Phys.\ Lett.\ }}

\def\ZP{{ Z.\ Phys.\ }}

\begin{document}

\hfill BI-TP 2004/01 

\vskip 2cm

\centerline{\LARGE \bf The Hagedorn Temperature}

\bigskip

\centerline{\LARGE \bf and Partition Thermodynamics}

\vskip 1.3cm 

\centerline{\large \bf Ph.\ Blanchard$^a$, S.\ Fortunato$^a$ 
and H.\ Satz$^{a,b}$} 

\vskip 0.7cm

\medskip

\centerline{a: Fakult\"at fur Physik, Universit\"at Bielefeld}

\centerline{Postfach 100 131, D-33501 Bielefeld, Germany}
 
\medskip

\centerline{b: Centro de F{\' \i}sica das Interac{\c c}{\~o}es 
Fundamentais (CFIF)}

\centerline{Instituto Superior T\'ecnico, Av. Rovisco Pais, P-1049-001 Lisbon, 
Portugal}

\vskip 1cm

\centerline{\bf \large Dedicated to Rolf Hagedorn, 1919 - 2003}

\vskip 1cm

\centerline{\bf Abstract:}

\bigskip

We review the resonance gas formalism of hadron thermodynamics
and recall that an exponential increase of the resonance spectrum leads
to a limiting temperature of hadronic matter. We then show that the number 
$p(n)$ of ordered partitions of an integer $n$ grows exponentially with
$n$ and satisfies the integer counterpart of the statistical bootstrap
equation. Considering the set of all partitions as a Gibbs ensemble 
provides a partition thermodynamics which is also governed by a limiting 
temperature, determined by the combinatorial structure of the 
problem. Further associating intrinsic quantum numbers to integers
results in a phase diagram equivalent to that found in QCD for hadronic 
matter as function of temperature and baryochemical potential. 
 
\bigskip

\newpage

\noindent
{\large \bf 1.\ Introduction}

\bigskip

The most important observation in particle physics around 1960 was 
that the number of different species of so-called elementary particles 
seemed to grow without limit. Hadron-hadron collisions produced more
and more resonant hadronic states of increasing masses. This phenomenon 
triggered two different theoretical approaches. 

\medskip

The more conventional idea, classical reductionism as pursued
in atomism since antiquity, proposed that there must be a smaller
number of more elementary objects, which then bind to form the 
observed hadrons as composite states. This approach, as we know,
ultimately led to quantum chromodynamics as the fundamental theory
of strong interactions and thus once more proved most successful. 

\medskip

A second, truly novel approach asked what such an increase of
states would lead to in the thermodynamics of strongly interacting
matter \cite{Hagedorn}. Both the question and the answer: the existence 
of an ultimate temperature of hadronic matter, are the contribution of
Rolf Hagedorn. We know today that strong interaction thermodynamics 
leads to critical behavior, to a phase transition in which hadronic
matter turns into a plasma of deconfined quarks and gluons.
Statistical QCD has confirmed this and led to a detailed picture
of this novel phenomenon. Let us see how it comes about and
what is its hadronic basis.

\vskip0.5cm

\noindent
{\large \bf 2.\  Hadron Thermodynamics}

\bigskip

Consider an ideal gas of identical neutral scalar particles of mass
$m_0$ contained in a box of volume $V$, assuming Boltzmann statistics.
The grand canonical partition function of this system is given by

\be
{\cal Z}(T,V) = \sum_N {1 \over N!} \left[ {V \over (2\pi)^3} \int d^3p~
\exp\{-\sqrt{p^2+m_0^2}~/T\} \right]^N, 
\ee

\noindent
leading to 

\be
\ln {\cal Z}(T,V) = {VTm_0^2 \over 2\pi^2}~ K_2({m_0\over T}).
\ee

\medskip

\noindent
For temperatures $T\gg m_0$, the energy density of the system becomes

\be
\e(T) = - {1 \over V}~ {\partial~\!\ln~\!{\cal Z}(T,V) \over \partial~\!(1/T)} 
\simeq {3 \over \pi^2}~ T^4,
\ee

\medskip

\noindent
the particle density

\be
n(T) =  {\partial~\!\ln~\!{\cal Z}(T,V) \over \partial~\!V}
\simeq {1 \over \pi^2}~ T^3
\ee

\bigskip

\noindent
and the average energy per particle

\be
\omega \simeq 3~T.
\ee

\newpage

\noindent
Hence an increase of energy of the system has three consequences:
it leads to

\begin{itemize}
\vspace*{-0.3cm}
\item{a higher temperature,}
\vspace*{-0.3cm}
\item{more constituents, and}
\vspace*{-0.3cm}
\item{more energetic constituents.}
\vspace*{-0.3cm}
\end{itemize}
If we now consider an interacting gas of basic hadrons and 
introduce resonance formation as the fundamental feature of 
their dynamics, we can approximate the interacting medium as
an ideal gas of the possible resonance species \cite{B-U}.
The partition function of this resonance gas is

\be
\ln {\cal Z}(T,V) = \sum_i
{VTm_i^2 \over 2\pi^2}~\rho(m_i)~ K_2({m_i\over T})
\label{resgas}
\ee

\medskip

\noindent
where the sum begins with the stable ground state $m_0$ and
then includes the possible resonances $m_i, i=1,2,...$ with
weights $\rho(m_i)$ relative to $m_0$. 
Clearly the crucial question here is how to specify $\rho(m_i)$,
how many states there are of mass $m_i$. It is only at this point
that hadron dynamics enters.

\medskip

The simplest answer is that $\rho(m)$ is obtained just from combinatorics, 
and it immediately gives us Hagedorn's statistical bootstrap model. 
Hagedorn assumed that ``fireballs consist of fireballs, which consist
of fireballs...''. In a more modern form, one would assume that resonance
formation and decay follow a self-similar pattern. In any case, the
defining equation for $\rho(m)$ is
\vskip-0.2cm
$$
\rho(m,V_0) = \delta(m\!-\!m_0) ~+
$$
\be
\sum_N {1\over N!} \left[ {V_0 \over (2\pi)^3} \right]^{N-1} 
\hskip-0.2cm \int \prod_{i=1}^N ~[dm_i~ \rho(m_i)~ d^3p_i] 
~\delta^4(\Sigma_i p_i - p).
\label{bootstrap}
\ee

\medskip

\noindent
It was solved analytically by W.\ Nahm \cite{Nahm}, giving
\be
\rho(m,V_0) = {\rm const.}~m^{-3} \exp\{m/T_H\}.
\label{exp}
\ee

\medskip

\noindent
The density of states thus increases exponentially in $m$, with
a coefficient $T_H^{-1}$ determined by
\be
{V_0 T_H^3 \over 2 \pi^2} (m_0/T_H)^2 K_2(m_0/T_H) = 2 \ln 2 - 1,
\label{bs}
\ee

\medskip

\noindent
in terms of two parameters $V_0$ and $m_0$. Hagedorn assumed that
the composition volume $V_0$, specifying the intrinsic range of 
strong interactions, is determined by the inverse pion mass as
scale, $V_0 \simeq (4 \pi/3) m_{\pi}^{-3}$. This leads to a temperature
$T_H \simeq 150$ MeV. It should be emphasized, however, that this is
just one possible way to proceed. In the limit $m_0 \to 0$, eq.\ (\ref{bs})
gives 
\be
T_H = [\pi^2 (2\ln 2 -1)]^{1/3}~ V_0^{-1/3} \simeq 1/r_h,
\ee
where $V_0 = (4\pi /3) r_h^3$ and $r_h$ denotes the range of
strong interactions. With $r_h \simeq 1$ fm, we thus have 
$T_H \simeq$ 200 MeV. From this it is evident that the
Hagedorn temperature persists in the chiral limit and is in fact
only weakly dependent on $m_{\pi}$, provided the strong interaction
scale $V_0$ is kept fixed. 

\medskip

If we now replace the sum in the resonance gas partition function 
(\ref{resgas}) by an integral and insert the exponentially growing mass 
spectrum (\ref{exp}),
$$
\ln {\cal Z}(T,V) \simeq 
{V T \over 2\pi^2} \int dm~m^2 \rho(m_i)~ K_2({m_i\over T})
$$
\be
\sim V\left[{T \over 2\pi}\right]^{3/2} \int dm~m^{-{3/2}} 
\exp\{-m \left[{1\over T} - {1\over T_H}\right]\},
\label{div}
\ee

\medskip

\noindent
we obtain a divergence for all $T > T_H$: in other words,
$T_H$ is the ultimate temperature of hadronic matter. In contrast to
what we found above, an increase of energy now leads to

\begin{itemize}
\vspace*{-0.2cm}
\item{a fixed temperature limit, $T \to T_H$,}
\vspace*{-0.3cm}
\item{the momenta of the constituents do not continue to increase, and}
\vspace*{-0.3cm}
\item{more and more species of ever heavier particles appear.}
\vspace*{-0.2cm}
\end{itemize}
We thus obtain a new, non-kinetic way to use energy, increasing the 
number of species, not the momentum per particle.

\medskip

Hagedorn originally interpreted $T_H$ as the highest possible 
temperature of strongly interacting matter. Somewhat later Cabibbo
and Parisi \cite{C-P} pointed out that the resonance gas partition function
(\ref{div}) in fact does not diverge at $T=T_H$, while its higher
derivatives do. Such behavior occurs at phase transition points, and
so $T_H$ is ``only" a critical temperature of strongly
interacting matter. It is clear now that $T_H$ indeed defines the
transition from hadronic matter to a quark-gluon plasma. 
Hadron physics alone can only specify its inherent limit; to go beyond 
this limit, we need QCD.

\medskip

The crucial feature leading to the observed limit of hadron physics
at $T_H$ is the exponential increase in the number of hadronic states.
In the following section, we want to study the origin of such an
increase and look somewhat closer at the nature of $T_H$.

\vskip0.5cm

\noindent
{\large \bf 3.\  Partition Thermodynamics}

\bigskip

To arrive at the simplest possible problem leading to an exponentially
increasing number of `states', we consider the number $p(n)$ of ordered
partitions of an integer $n$ into integers. To illustrate: for $n=3$,
we have the ordered partitions 3, 2+1, 1+2, 1+1+1, so that
\be
p(n=3) = 4 = 2^{n-1}.
\ee
Similarly, $n=4$ gives the partitions 4, 3+1, 2+2, 2+1+1, 1+3, 1+2+1, 
1+1+2, 1+1+1+1, leading to
\be
p(n=4) = 8 = 2^{n-1}.
\ee
This solution can be shown to hold in fact for all $n$ (see Appendix),
i.e.,
\be
p(n) = 2^{n-1}= {1 \over 2} ~\exp\{n \ln 2\}.
\label{part1}
\ee
We note here for completeness that calculating the number $q(n)$ of 
unordered partitions of an integer $n$ (i.e., not counting permutations) 
is more difficult and solved only asymptotically \cite{Hardy}:  
\be
q(n) =
{1\over 4 \sqrt 3~ n} e^{\{\pi \sqrt{2n/3}\}}\Big[1+O(\frac{\,\log\,n}{n^{1/4}})\Big]
\ee
We will not make use of this result here.

\medskip

The problem of ordered partitions can also be solved for restricted
or extended cases. For example, if we allow only odd integers in the
partitions, we get for large $n$
\be
p(n) \simeq {\frac{\kappa}{\sqrt{5}}}\,e^{n\,\ln\,\kappa} 
\ee
with $\kappa = (1+\sqrt 5)/2 \simeq 1.62 < 2$. On the other hand, if
we introduce further degrees of freedom, e.g., giving each integer 
an intrinsic quantum number (`spin') which can take on the values
$\pm 1$, the exact solution is
\be
p(n) = 2 \kappa^{n-1} = {2\over \kappa}~\!e^{\{n \ln \kappa\}},
\label{part2}
\ee
now with $\kappa = 3 > 2$. Hence we have as solution to
the generalized ordered partition problem the form (see Appendix)
\be
p(n) \propto {\kappa}^n = e^{\{n \ln \kappa\}}
\label{part3}
\ee     
with $\kappa = 2$ for the ``standard" case and in general
$\infty > \kappa > 1$.

\medskip

We thus obtain an exponential increase in the number of possible
partitions. How is this related to the statistical bootstrap condition
(\ref{bootstrap})? Consider a density of states defined by a bootstrap
equation for integers,
\be
\rho(n) = \delta(n\!-\!1) + \sum_{k=2}^n~ {1 \over k!}~ \prod_{i=1}^k
~\rho(n_i)~ \delta(\Sigma_i n_i\! -\!n).
\label{nbootstrap}
\ee
Its solution is just the number of partitions of $n$,
\be
\rho(n) = z~p(n)
\ee
up to a normalization of order unity (for the standard case $\kappa=2$,
$z \simeq 1.25$). We can thus conclude

\begin{itemize}
\vspace*{-0.2cm}
\item{the statistical bootstrap equation is an integral formulation
of a partition problem;}
\vspace*{-0.3cm}
\item{the exponential increase in the number of states is of combinatoric
origin, and}
\vspace*{-0.3cm}
\item{the exponential coefficient $\kappa$ is determined by the
combinatoric structure.}
\vspace*{-0.2cm}
\end{itemize}
Let us therefore consider $\kappa$ in more detail. In the 
standard case, see eq.\ (\ref{part1}), the exponential increase in 
the number of partitions is determined by $\ln 2$; in eq.\ (\ref{bs}),
we had seen that in the actual statistical bootstrap, $T_H$ is also 
determined by $\ln 2$, but in a somewhat more complex way due to the 
presence of momentum degrees of freedom in addition to the hadron masses.

\medskip

To understand the meaning of $\kappa$ in a thermal context,
we  construct a statistical mechanics of partitions,
considering each partition as a point in the `phase space' of all 
partitions of fixed $n$. The set of all partitions thus forms a 
Gibbs ensemble, and we define the partition entropy as
\be
S(n) = \ln p(n) \simeq n \ln \kappa,
\label{entropy} 
\ee
the specific partition entropy as
\be
s(n) = {S(n) \over n} \simeq \ln \kappa
\label{sentropy}
\ee
and the partition temperature $\Theta$ as 
\be
 {1 \over \Theta} = {dS(n) \over dn} = \ln \kappa.
\label{temp}
\ee
The parameter $\ln \kappa$ determining the increase of the number of 
partitions of $n$ is thus the partition temperature $\Theta$, which 
here for large $n$ coincides with the inverse of the
specific partition entropy. 

\medskip

Next we turn to the mentioned generalized form, in which each integer
$n_i$ has an intrinsic `spin' variable $\sigma_i = \pm 1$, in order
to see what the effect of conserved quantum numbers is. We define the
total spin for a given partition of $n$ as
\be
\sigma_T(n) = \sum_i \sigma_i,
\ee
where the sum runs over all $k\leq n$ terms of the partitions. We now 
calculate the number $p(n,\sigma_T)$ of partitions of $n$ at fixed
$\sigma_T$. For $\sigma_T=n$ we have $p(n,n) = 1$, since  
only the partition $n=1+1+...+1$ of all spins $+1$ can give a total
spin equal to $n$. For $\sigma_T=0$ we find $p(n,0) \simeq 3^{n-1}$,
since total spin zero constitutes the dominant part of all configurations. 
The specific partition entropy $s(n,\sigma_T)$ is shown in Fig.\ \ref{sent}.
It was calculated numerically
for values of $n$ up to $20000$ (see Appendix) and it was 
found to depend only on $x=\sigma_T/n$. Moreover, our calculations show 
no appreciable difference between $n=10000$ and $n=20000$, so that the
latter is in good approximation the asymptotic form of the curve. 
It can be fitted by 
\be
s(x) = \left[ 1 - x^a\right]^b \ln 3,
\label{fit}
\ee
with a=2.1, b=0.736, and is shown in Fig.\ \ref{sent}, varying from 
$s(0)=\ln 3$ to $s(1)=0$. 

\begin{figure}[h]
\centerline{\psfig{file=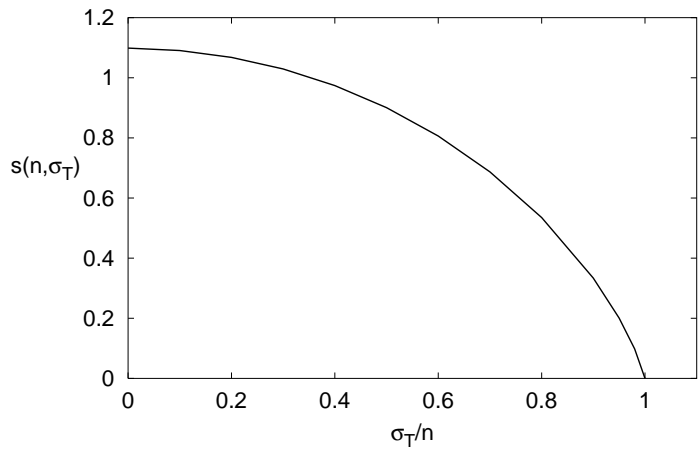,width=9cm}}
\vspace*{-0.3cm}
\caption{Specific partition entropy as function of $x=\sigma_T/n$}
\label{sent}
\end{figure}

\medskip

\begin{figure}[h]
\centerline{\psfig{file=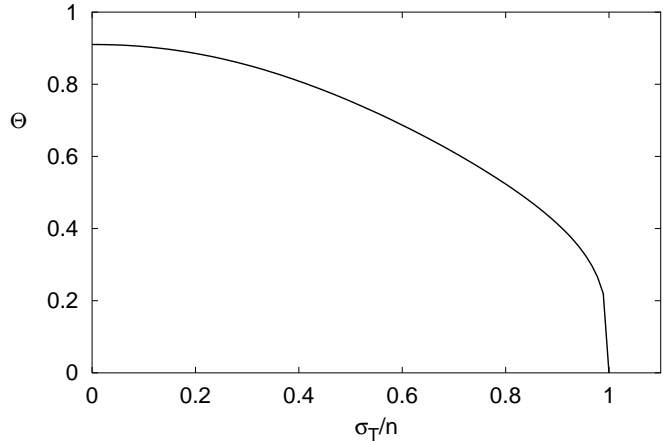,width=9cm}}
\vspace*{-0.3cm}
\caption{Partition temperature as function of $x=\sigma_T/n$}
\label{thetax}
\end{figure}

The partition temperature now becomes
\be
{1\over \Theta(x)} = {dS(n,\sigma_T) \over dn} = 
{d[n\,s(x)] \over d n} = 
s(x) - x {d s(x)
\over d\,x}
\label{iox}
\ee
and is thus no longer equivalent to the inverse specific partition
entropy: for $x \to 1$, both $s(x)$ and $\Theta(x)$ vanish. Through 
(\ref{iox}) we can determine $\Theta(x)$ from the function
$s(x)$ (see eq.\ (\ref{fit})) that we have previously calculated. The
resulting form is shown in Fig.\ \ref{thetax}; it decreases from
$\Theta = 1/\ln 3$ at $x=0$ to 0 at $x=1$.

The constituents of our partition thermodynamics are
the different integers occurring in the partitions; they are the
resonances of the resonance gas. It is thus interesting to ask for
the relative abundance $R_{n,\sigma_T}(a,b)$ of two integers $a,b < n$ in
the set of all partitions of $n$. In the appendix, we calculate
the total occurrence $h(n,a, \sigma_T)$ of an integer $a$ in the 
partitions of $n$ at fixed $\sigma_T$. Given this, the desired
relative abundance is defined as 
\be
R_{n,\sigma_T}(a,b) = {h(n,a,\sigma_T) \over h(n,b,\sigma_T)}
\ee
For $a,b \ll n$ it turns out that for large $n$, $R_{n,\sigma_T}(a,b)$ 
depends only on the ratio $\sigma_T/n$ and has the form
\be
R_{n,\sigma_T}(a,b) \simeq e^{-(a-b)/\Theta(\sigma_T/n)},
\ee
where $\Theta$ is precisely
the partition temperature shown in Fig. \ref{thetax}. 
We thus see that the resonance gas feature of species 
abundances specified by one universal temperature in fact has its
origin in the underlying combinatoric structure.

\medskip

\begin{figure}[h]
\centerline{\psfig{file=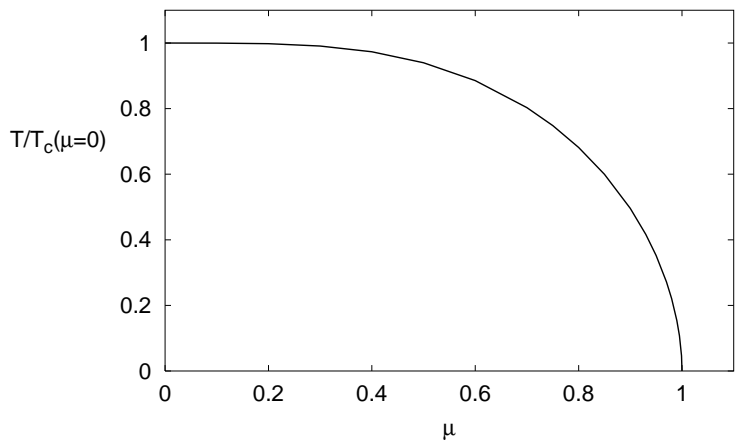,width=9cm}}
\vspace*{-0.3cm}
\caption{Phase boundary for partition thermodynamics}
\label{diagram}
\end{figure}

So far, we have considered partition thermodynamics in what might
be called a grand microcanonical formulation: the number of terms
in the different partitions varies, but their sum $n$ and the overall
spin $\sigma_T$ are specified exactly. To go to a grand canonical
formulation, we fix $n$ and $\sigma_T$ only on the average. The
starting point is the grand canonical partition function
\be 
{\cal Z}(T,\mu)\! = \! \int_1^{\infty}\! dn \exp\{-n/T\}\! 
\int_{-n}^{n}\! d\sigma_T 
\exp\{-\mu \sigma_T/T\}
~p(n,\sigma_T),
\label{3.1}
\ee 
with the temperature $T$ and the chemical potential $\mu$ the
Lagrangian multipliers for $n$ and $\sigma_T$. 
Since $p(n,\sigma_T)=\exp\{ns(\sigma_T/n)\}$, where $s(\sigma_T/n)$ 
is the specific entropy calculated above, we get
\be
{\cal Z}(T,\mu)=\int_1^{\infty}\,dn \,\int_{-n}^n
d\sigma_T\,\exp\{-n\left[\frac{1}{T}-
\frac{\mu}{T}\frac{\sigma_T}{n}-s(\sigma_T/n)\right]\}.
\label{3.2}
\ee
\noindent
The integral (\ref{3.2}) is not defined for all pairs of values 
$(T,\mu)$:  for fixed $\mu$,
it exists only for those values of the temperature for which
\be
\frac{1}{T}-\frac{\mu}{T}\frac{\sigma_T}{n}-s(\sigma_T/n)>0\,
~~\Rightarrow~~
T<\frac{1-\mu x}{s(x)} \hskip1cm \forall\, \sigma_T,n.
\label{3.3}
\ee
\noindent
We note that the function on the right of the last inequality depends 
only on the variable $x=\sigma_T/n$, and not on $\sigma_T$ and $n$
separately.
Since the condition (\ref{3.3}) must be valid for any possible value of the
ratio $\sigma_T/n$, with $|\sigma_T/n|\leq\,1$, 
$T$ has to be smaller than the absolute minimum of the function
\be
f(x,\mu)=\frac{1-\mu\,x}{s(x)},
\label{3.4}
\ee
in the interval $-1\leq x \leq 1$.
Hence for any $\mu$ there exists a limiting value $T_c(\mu)$ of the 
temperature beyond which the grandcanonical partition function is not defined.
In particular, for $\mu=0$, we get $T_c=1/\ln\,3$, which is the temperature
determined in the previous 
section for the partition problem with total spin $\sigma_T=0$.
We further note that $|\mu|\leq\,1$, because otherwise the function
$f(x,\mu)$ would become negative for some $x$ and the inequality (\ref{3.3})
could never be satisfied by positive values of the temperature $T$.   
By varying $\mu$ we obtain a curve in the $T-\mu$ plane,
which defines the existence region for our system.
Determining numerically the minimum of the $f(x,\mu)$ as function of $\mu$,
we obtain the form shown in Fig. \ref{diagram}.

\medskip

The resulting `phase diagram' 
of partition thermodynamics, i.e., the boundary for the existence of such
thermodynamic systems, is evidently very similar to that expected in
statistical QCD for the phase boundary of hadronic matter. It thus
seems that quantum chromodynamics leads to a resonance pattern which
is basically of combinatoric nature. Moreover, the conservation of
baryon number also appears to follow a simple combinatoric form
resulting from the addition of the baryonic degrees in the resonance
composition.

\medskip

\noindent
{\large \bf 4.\  Conclusions}

\bigskip

We have shown that the limit of hadron physics obtained in resonance gas 
thermodynamics with an exponentially increasing mass spectrum is due to
the combinatorics of resonance formation. In particular, the simplest
partition systems provide 

\begin{itemize}
\vspace*{-0.2cm}
\item{a limiting temperature determined solely by the combinatoric structure
of the system;}
\vspace*{-0.3cm}
\item{a reduction of this limiting temperature for conserved intrinsic
quantum numbers and hence a phase boundary for partition thermodynamics;}
\vspace*{-0.3cm}
\item{a relative abundance of constituents governed by the limiting
temperature.}
\vspace*{-0.2cm}
\end{itemize}

\noindent
Hence, if all combinations of $n$ hadrons are resonances and the 
degeneracy of each resonance is determined by the number of composition
patterns, an energy input eventually goes into species formation,
not into faster constituents. Comparing such a resonance gas to a
conventional pion gas, we thus find for the dependence on energy density
$\e$ 

\bigskip

\hskip 2cm pion gas \hskip 6cm resonance gas

\medskip

\hskip 2cm
$ n_{\pi} \sim  \e^{3/4} \hskip 6cm n_{res} \sim \e$ 

\hskip 2cm
$ \omega_{\pi} \sim \e^{1/4} \hskip 6cm \omega_{res} \sim {\rm const.}$

\bigskip

\noindent
Here $n$ denotes the density of constituents, $\omega$ the average
energy of a constituent. 

\bigskip

We close with a rather speculative question, concerning the relation of 
our considerations to nuclear collisions. In the experimental study of 
such interactions, it is found that the relative abundances of hadron 
species are essentially determined by two parameters, the temperature $T$ 
of a resonance gas and the baryochemical potential $\mu$ specifying its
overall baryon density \cite{Beca}. For high energy collisions at $\mu=0$, 
$T\simeq 175$ MeV; with increasing $\mu$, $T(\mu)$ decreases in a
form quite similar to that shown in Fig.\ \ref{diagram}. What can we
learn from this? There seem to be two possible scenarios:

\begin{itemize}
\vspace*{-0.2cm}
\item{Each collision produces a thermal system and thus corresponds to 
an ensemble of many partitions: nuclear collisions produce matter.}
\vspace*{-0.3cm}
\item{Each collision is one partition, and only the ensemble over
many collisions forms a thermal ensemble: nuclear collisions simulate
matter.}
\vspace*{-0.2cm}
\end{itemize}
  
\noindent
Which of these two is realized in nature appears to be one of the
most crucial questions for the experimental high energy heavy ion
program. 

\bigskip

\noindent
{\large \bf Acknowledgements}

\medskip

S.F.\ acknowledges the support of the DFG Forschergruppe
under grant FOR 339/2-1.

\bigskip
~

\noindent
{\large \bf Appendix}

\bigskip

\noindent
{\bf A.\ Standard Case}

\medskip

To determine the number of ordered partitions of an integer $n$ into 
integers $k_i \leq n$, we start from the number of partitions containing
$m$ terms. This is equivalent to the number of divisions of a segment 
of length $n$ units into $m$ parts. 
Each partition of this type is constructed by choosing $m-1$
of the first $n-1$ integers as end points for the $m$ segments
dividing $[0,n]$. This can be done in 

\be
p(n,m)={n-1\choose m-1}
\label{op1}
\ee

\medskip

\noindent 
possible ways. By summing over all possible values of $m$, we obtain 

\be
p(n)=\sum_{m=1}^{n}p(n,m)=\sum_{m=1}^{n}{n-1\choose m-1}=2^{n-1}.
\label{op2}
\ee

\medskip

\noindent 
Next, we determine the abundance of a given integer $a$ in the set of  
all partitions. Suppose we have a partition of $m$ terms with at least 
one occurrence of the number $a<n$. By changing the position of $a$ within 
the partition, we get $m$ possible partitions containing the same number
of terms, with order of the $m-1$ other terms unchanged. Hence the number 
of occurrences of $a$ in partitions with $m$ summands is just $m$ times 
the number of partitions of $n-a$ into $m-1$ terms. Among the partitions 
of $n-a$ into $m-1$ terms there will in general be other $a$'s, and in 
this case each partition gives a  contribution larger than one to the 
occurrence of $a$. However, this larger contribution is exactly compensated 
by a reduction of the number of corresponding partitions, since
the different $a$ are indistinguishable and can be freely permuted 
without generating new partitions. To get the total occurrence $h(n,a)$ of 
$a$ in all partitions of $n$ we sum over all possible values of $m$ and
obtain

\be
h(n,a)=\sum_{m=2}^{n-a+1}m\,{n-a-1\choose m-2}.
\label{op3}
\ee

\medskip

\noindent
Defining $l=m-2$, we get

\be
h(n,a)=\!\!\sum_{l=0}^{n-a-1}(l+2)\,{n-a-1\choose l}=2\,
\!\!\sum_{l=0}^{n-a-1}\,{n-a-1\choose
l}+\!\!\sum_{l=0}^{n-a-1}l\,{n-a-1\choose l}.\!
\label{op4}
\ee

\medskip

\noindent
and from this

\be
h(n,a)=(n-a-1)\,2^{n-a-2}+2^{n-a}=\frac{n-a+3}{4}\,2^{n-a}
\label{op5}
\ee

\medskip

\noindent
for the abundance of $a$ in the partitions of $n$.

\bigskip

\noindent
{\bf B.\ Intrinsic Degrees of Freedom}

\medskip

We now assume that each integer carries an intrinsic degree of freedom, 
an integer `spin' $\sigma$, giving it $2\sigma$ degenerate states; in the
example above, we had $\sigma=1$ and hence the two states $\pm\,1$.
It is obvious that such a spin leads to an increase in the number of
partitions; each partition is now labeled by $n$ and by the total spin 
$\sigma_T=\sum_{k}\sigma_k$ of its summands. If we do not impose any 
constraint on the total spin, it is trivial to calculate the total number 
of partitions. There are now $2\sigma$ copies of each number, so that in 
a partition with $m$ terms, each summand contributes an additional factor 
$2\sigma$. Eq. (\ref{op1}) thus becomes 
\be
p_{\sigma}(n,m)=(2\sigma)^m\,{n-1\choose m-1}
\label{op6}
\ee
\noindent so that
\be
p_{\sigma}(n)=\sum_{m=1}^{n}p_{\sigma}(n,m)
=\sum_{m=1}^{n}(2\sigma)^m\,{n-1\choose m-1}=(2\sigma)\,(2\sigma+1)^{n-1}.
\label{op7}
\ee
The problem is more interesting if $\sigma_T$ has a well defined value,
which is equivalent to requiring the conservation of a spin, isospin
or baryon number for a system of particles. In this case, it is not possible 
to compute the total number of partitions and the abundances in closed form. 
However, at least for the special case $\sigma=1$, we can derive formulas 
that allow a numerical computation of the quantities of interest. Since the 
problem is completely symmetric under a change of sign of the spin, we
consider positive values of the total spin.

We start by asking how many partitions of $m$ terms there are with a total 
spin $\sigma_T$. The summands of the partitions carry spin $+1$ or
$-1$. If there are $k$ numbers with spin $+1$ there must then be $m-k$ 
numbers with spin $-1$. In order to have a total spin $\sigma_T$, there is 
only one possibility: $(m+\sigma_T)/2$ terms must carry spin $+1$ and 
$(m-\sigma_T)/2$ terms spin $-1$. Hence the required number of partitions 
at fixed $m$ is
\be
p(n,m,\sigma_T)={m\choose (m+\sigma_T)/2}{n-1\choose m-1}
\label{op8}
\ee

\noindent and the total number becomes
\be
p(n,\sigma_T)=\sum_{m=\sigma_T}^{n}p(n,m,\sigma_T)=
\sum_{m=\sigma_T}^{n}{m\choose (m+\sigma_T)/2}{n-1\choose m-1}.
\label{op9}
\ee
Note that the sum over $m$ is constrained by the parity of the total spin:
if $\sigma_T$ is even (odd), $m$ must be even (odd). Hence the sums run over
those values of $m$ between $\sigma_T$ and $n$ which have the same parity 
as $\sigma_T$.

To obtain the abundances, we only have to insert in Eq. (\ref{op3}) the 
factor due to the spin degeneracy of the partitions, given in Eq. (\ref{op9}), 
leading to 
 \be
h(n,a,\sigma_T)=\sum_{m=\sigma_T}^{n}m\,{m\choose (m+\sigma_T)/2}{n-a-1\choose m-2}.
\label{op10}
\ee
Eqs. (\ref{op9}) and (\ref{op10}) can easily be evaluated using 
{\it Mathematica}.

\end{document}